\documentclass[aps,prb,twocolumn,showpacs,amsmath,amssymb]{revtex4-1}
\usepackage[linktocpage=true]{hyperref}
\usepackage{graphicx}
\usepackage{booktabs}
\usepackage{epsfig}
\usepackage{color}
\usepackage{epstopdf}

\newcommand{\E}{\mathcal{E}}
\newcommand{\T}{\mathcal{T}}
\newcommand{\red}[1]{\textcolor{black}{#1}}

\begin{document}

   \title{
Energy and  power fluctuations in  ac-driven coherent conductors} 
         \author{Francesca Battista$^{1,2}$, Federica Haupt$^{1,3}$, and Janine Splettstoesser$^{1,4}$}
         \affiliation{$^1$Institut f\"ur Theorie der Statistischen Physik, RWTH Aachen University, D-52056 Aachen, Germany\\
          \& JARA-Fundamentals of Future Information Technology\\
$^2$Departamento de F\'isica, FCEyN, Universidad de Buenos Aires and IFIBA,
Pabell\'on I, Ciudad Universitaria, 1428 CABA Argentina\\
$^{3}$JARA Institute for Quantum Information, RWTH Aachen University, D-52056 Aachen, Germany\\
$^4$Department of Microtechnology and Nanoscience (MC2), Chalmers University of Technology, SE-41298 G\"oteborg, Sweden}
\date\today

\begin{abstract}
Using a scattering matrix approach we study transport in coherent conductors driven by a time-periodic bias voltage. We investigate the role of electron-like and hole-like excitations created by the driving in the energy current noise and we reconcile  previous studies on charge current noise in  this kind of systems. The energy  noise reveals \textit{additional} features due to electron-hole correlations. These features should be observable in power fluctuations. In particular, we show results for the case of a harmonic and bi-harmonic driving and of Lorentzian pulses applied to a two-terminal conductor, addressing the recent experiments of Refs. \onlinecite{Gabelli13,Dubois13nat}.
\end{abstract}
\pacs{72.70.+m, 73.23.-b, 85.36.Gv}
\maketitle

\section{Introduction} 

In recent years the interest in transport through mesoscopic systems driven by 
time-dependent potentials has been increasing, in particular with the aim of obtaining enhanced control over the charge flow through the system. Indeed, even the controlled emission of \textit{single electrons} (or a fixed integer number of electrons) in a given time interval has been realized with periodically and nonlinearly driven mesoscopic capacitors~\cite{Feve07,Hermelin11,McNeil11} and electron pumps.~\cite{Pekola13rev,Leek05,Blumenthal07,Kaestner08,Maisi09,Connolly13}
These setups are usually
based on the emission of particles from a confined region such as a quantum dot. They are useful for metrology, where a quantum standard for the current is sought for, or for the tuneable synchronization of particles for quantum optics with electrons.~\cite{Olkhovskaya08,Bocquillon12,Ubbelohde14} 
It is hence necessary to make these single-electron sources as precise as possible,~\cite{Keeling06,Moskalets08,Keeling08,Mahe10,Parmentier12}  decreasing their noise. Furthermore, it is important to achieve a profound understanding of the properties of the source,~\cite{Albert10} for example by explicitly studying the waiting time distribution of emitted particles~\cite{Tang14,Albert11,Dasenbrook14} and the energy spectrum of the signal.~\cite{Moskalets11, Battista12, Ferraro13, Fletcher13, Moskalets14}

A different way of creating a controlled and noiseless flow of single particles - without resorting to specific confined structures - is the application of Lorentzian voltage pulses to a conductor. It has been demonstrated by Levitov and coworkers~\cite{Ivanov97,Keeling06,Keeling08} that a series of Lorentzian-shaped pulses $V(t)$ of quantised flux, $\int eV(t)dt=h$, 
leads to the propagation of a noiseless train of independent \textit{single-particle} excitations, which were therefore named \textit{Levitons}. This means that, in striking contrast with the general situation where  an oscillating voltage is applied, no electron-hole pairs induced by this particular time-dependent driving contribute to  transport. Recent experiments showed that by superposing several harmonic driving potentials a Lorentzian pulse carrying an integer number of particles can
be  approximately  achieved, leading to a reduction of the charge current noise reduction~\cite{Gabelli13} and allowing for the study of controlled single-particle effects.~\cite{Dubois13nat} 

These recent efforts have boosted the general interest in the study of  noise~\cite{Blanter00} as a powerful tool to characterize the response of a conductor to a generic time-dependent driving potential.~\cite{Vanevic12,Parmentier12,Jonckheere12,Zamoum12,Gasse13,Gabelli13,Dubois13,Dubois13nat,Moskalets13} In particular it has been shown that the noise carries information on the probability with which electron-hole pairs are created by the ac part of the driving~\cite{Vanevic07,Vanevic08,Vanevic12} and it is sensitive to correlations between electrons and holes.~\cite{Rychkov05} Electron-hole pairs do not carry charge and therefore their creation by the ac potential does not affect the average charge transport. However, these pairs carry a finite energy and they will thus  strongly influence the properties of energy transport in  ac-driven conductors.

Here, we study a simple system consisting of a two-terminal conductor with a central scatterer. This system is subjected to an arbitrary, time-dependent
bias. Its study allows for general statements concerning the impact of electron and hole excitations on charge and energy transport and their fluctuations and it is at the same time appropriate to model the latest experiments on controlled charge transport.~\cite{Gabelli13,Dubois13nat}
After revisiting the study of the charge current noise in ac-driven systems,~\cite{Schoelkopf98,Reydellet03,Rychkov05,Moskalets04,Moskalets06,Vanevic07,Keeling08,Vanevic08,Andrieux09,Vanevic12,Gasse13,Gabelli13}  we go beyond this study by considering in detail also the energy current and energy current fluctuations. 

Our analysis of the charge current noise considers in detail contributions due to \textit{correlations between electrons and holes} created by the driving potential as well as the \textit{separate contributions} of electron-like and hole-like excitations, and reconciles interpretations obtained from previous works~\cite{Rychkov05,Reydellet03,Vanevic07,Vanevic08,Vanevic12} where the impact of electron-hole pairs and their correlations was debated. 
It turns out to be useful to separate the charge current noise into a \textit{transport} part and an \textit{interference} part.
We then consider in the same terms the energy current and energy current fluctuations.  
Motivated by the experimental progress in time-dependently driven electronic systems, there has indeed recently been growing interest in various aspects of their energy-transport properties.~\cite{Moskalets11,Battista12,Phan13,Gasparinetti14,Moskalets14} In this work, we show that while the energy flow can be interpreted as the time-average of the energy current due to a ``frozen" potential, energy fluctuations show specific features of the excitations created by the ac driving. In particular, we reveal features due to electron-hole correlations which are not visible in the charge noise and analyse their behavior for different types and superpositions of driving potentials, namely a harmonic or bi-harmonic driving as well as Lorentzian pulses. Specifically, for a system driven by Lorentzian-shaped pulses, we show that the energy and its fluctuations are a tool to reveal that $L$ particles emitted by a sequence of $L$ Lorentzians, each having a time integral equal to $h/e$, are independent, while $L$ particles emitted at once by a Lorentzian with a time integral equal to $L\times h/e$ are not~\cite{Ivanov97} - a characteristic which is not visible neither in the charge current nor in its noise.

All the major features discussed in the energy noise can be shown to be present also in the power fluctuations. They are hence expected to be measurable in up-to-date experimental setups.\cite{Gabelli13,Dubois13nat}

The manuscript is organized as follows: in Section~\ref{sec:model} we introduce the Floquet scattering matrix approach for time-dependent systems that we apply here; more details are given in the Appendix. The appearance of electron-like and hole-like excitations in charge and energy currents and their fluctuations are calculated in Section~\ref{sec:ehpicture}, followed by an interpretation of their contributions to the transport and interference parts of the noise, Section~\ref{sec:driving}. Finally, we relate the energy noise to the heat noise and to measurable power fluctuations in Section~\ref{sec:power}.

\section{Formalism}\label{sec:model}

We consider a coherent mesoscopic conductor connected to metallic contacts (also called reservoirs). The system is brought out of equilibrium by time-periodic voltages $V_{\alpha}(t)$ applied to these contacts. We describe charge and energy transport through the system  within the scattering theory of photon-assisted tunneling developed by B\"uttiker and coworkers,\cite{Pretre96,Pedersen98}  a brief summary of which is given in the Appendix. As customary, we model the conductor in terms of a central scattering region connected to the external electronic reservoirs by ideal leads, i.e. pieces of ballistic conductors. The total charge  current operator in contact $\alpha$ is~\cite{Buttiker92}  
\begin{equation}\label{eq:current}
\hat{I}_{\alpha}(t)=\frac{e}{h}  \int^{\infty}_{-\infty} dE\, dE' e^{i(E-E')t/\hbar}\ \hat{i}_{\alpha}(E,E'),
\end{equation} 
where  $e$ is the electron charge ($e<0$),  $\hat{i}_{\alpha}(E,E')=\hat{\bf b}^{\dagger}_{\alpha}(E)\hat{\bf b}_{\alpha}(E')-\hat{\bf a}^{\dagger}_{\alpha}(E)\hat{\bf a}_{\alpha}(E')$, and  $\hat{\bf a}_{\alpha}$ and  $\hat{\bf b}_{\alpha}$ are vectors of operators with components $\hat{a}_{\alpha n}$ and $\hat{b}_{\alpha n}$. The operator $\hat{ a}_{\alpha n}$ ($\hat{ b}_{\alpha n}$) annihilates  an electron impinging on (outgoing from) the scatterer in channel $n$ in lead $\alpha$. 
The relation between the operators $\hat{\bf a}_{\alpha}$ and  $\hat{\bf b}_{\alpha}$
is governed by the time-independent scattering matrix of the conductor,  $\hat{\bf b}_{\alpha}(E)=\sum_{\beta}{\bf s}_{\alpha \beta}(E)\hat{\bf a}_{\beta}(E),$ where ${\bf s}_{\alpha \beta}(E)$ has dimensions $N_{\alpha}\times N_{\beta}$ for leads with $N_{\alpha}$ and $N_{\beta}$ channels. Similarly, assuming that energy is carried only by electrons, the  energy flux in contact $\alpha$ is given by the operator\cite{LIm13,Ludovico13}  
\begin{equation}\label{eq:Ecurrent}
\hat{I}^{\mathcal{E}}_{\alpha}(t)
=\frac{1}{h} \int^{\infty}_{-\infty}\!\!  dE\,dE' \tfrac{(E+E')}{2}e^{i(E-E')t/\hbar}\hat{i}_{\alpha}(E,E'). 
\end{equation}

The physical observables that we are interested in are the dc-components of the average charge and energy currents 
\begin{gather}\label{eq:currents}
I_{\alpha}= \overline{\langle\hat{I}_{\alpha}(t)\rangle},\quad \quad \ I^{\E}_{\alpha}= \overline{\langle\hat{I}_{\alpha}^{\E}(t)\rangle},
\end{gather} 
as well as their zero-frequency auto- and cross-correlators 
\begin{subequations}\label{eq:noise-gen}
\begin{align}   
S_{\alpha\beta}&=\int_{0}^{\mathcal{T}}\frac{dt}{\mathcal{T}}\int_{-\infty}^{\infty}d\tau \langle \Delta \hat{I}_{\alpha}(t+\tau)\Delta \hat{I}_{\beta}(t) \rangle, \label{eq:S}\\
S_{\alpha\beta}^{\E}&= \int_{0}^{\mathcal{T}}\frac{dt}{\mathcal{T}}\int_{-\infty}^{\infty}d\tau \langle \Delta \hat{I}_{\alpha}^{\E}(t+\tau)\Delta \hat{I}_{\beta}^{\E}(t) \rangle, \label{eq:SE}
\end{align}
\end{subequations}
which in the following we will name in short the \textit{charge noise} and the {\em energy noise}, respectively.  Angular brackets denote the quantum and statistical average 
and $\Delta\hat{A}=\hat{A}-\langle \hat{A}\rangle$. The over-line indicates the time average over one period of the driving  $\mathcal{T}$, i.e.  $\overline{x(t)}= \int_{0}^{\mathcal{T}}\frac{dt}{\mathcal{T}} x(t) $. In addition to the charge and energy noise, also the mixed correlator between charge and energy currents is expected to be nonzero, since electrons and holes are carriers both of charge and energy. A discussion of this mixed correlator will be presented elsewhere.~\footnote{Note that for \textit{ideal} single-particle sources emitting electrons and holes to \textit{floating contacts}, the correlator between the induced fluctuations of the chemical potential and of the temperature in the contacts vanishes.~\cite{Battista13}}

In order to evaluate the quantum statistical averages, it is useful to
re-express the current operators in terms of the operators for incoming states $\hat{a}_{\alpha n}$ only.\cite{Buttiker92} In case of  time-independent voltages $V_\alpha$,  the statistics of these operators reflects directly the equilibrium statistical properties of the reservoirs.\cite{Buttiker92} Assuming that the latter are in thermal equilibrium, this implies 
$\langle \hat{a}_{\alpha n}^{\dag}(E) \hat{a}_{\beta m}(E')  \rangle=\delta_{\alpha\beta}\delta_{mn}\delta(E-E')f_{\alpha}(E)$, where $f_\alpha(E)=\left[1+\exp\{(E-eV_\alpha)/k_{\rm B}T\}\right]^{-1}$ is the Fermi function, with $T$ the electronic temperature in the reservoirs and $k_\mathrm{B}$ the Boltzmann constant.

This is no longer true in the presence of a time-dependent driving, because the ac part of the voltage gives rise to a spread in energy of the electronic wave-function. In this case, a state with energy $E$ in the leads corresponds to a superposition of  reservoir states with energy $E-k\hbar\Omega$, where $\Omega=2\pi/\T$ is the frequency of the driving.\cite{Pretre96,Pedersen98}  The statistics of the operators  $\hat{a}_{\alpha n}(E)$ and $\hat{a}_{\alpha n}^{\dag}(E)$ is thus modified into 
\begin{align}
\langle \hat{a}_{\alpha n}^{\dag}(E) \hat{a}_{\beta m}(E')  \rangle\!&=\!\delta_{\alpha\beta}\delta_{mn}\times \label{eq:contraction} \\
&\sum_{k,\ell=-\infty}^{+\infty}\!\!\! c^{*}_{\alpha k}c_{\beta k+\ell}f_{\alpha}(E_{-k})\delta(E_{\ell}\!-\!E')\nonumber
\end{align}
with $E_{\ell}=E+\ell\hbar \Omega$, and 
\begin{equation}\label{eq:ck}
c_{\alpha k}=\int_{0}^{\T} \frac{dt}{\T}\, e^{-i\frac{e}{\hbar}\int_{0}^{t} dt' [V_{\alpha}(t')-\overline{V}_{\alpha}]}\,e^{ik\Omega t}.
\end{equation}
Here $V_{\alpha}(t)$ is the voltage applied to contact $\alpha$ and  $\overline{V}_{\alpha}$ is its dc component. 
The coefficients $c_{\alpha k}$ represent the probability amplitude that an electron absorbs ($k>0$) or emits ($k<0$) $k$ energy quanta $\hbar\Omega$  (Floquet quanta) while interacting with the ac part of the driving voltage. Note that in Eq.~\eqref{eq:contraction}, the Fermi distribution depends only on the dc component $\overline{V}_{\alpha}$ of the potential applied to the contact.\cite{Pretre96,Pedersen98}

For the sake of clarity, from now on we will focus on the case of a two-terminal conductor, assume that the right contact is grounded $V_{\rm R}(t)=0$, and measure all energies with respect to the electrochemical potential $\mu$ of this reservoir. The left contact is subject to the time-dependent potential $V_{\rm L}(t)=V_{\rm ac}(t)+\bar{V}$, where $\bar{V}$ is the dc voltage offset and $V_{\rm ac} (t)$ is a pure ac component.

\section{Electron-hole picture of currents and fluctuations}\label{sec:ehpicture}

\begin{figure}[t,b]
\centerline{\psfig{figure=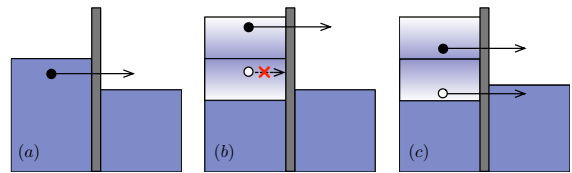,width=0.9\columnwidth}}
\caption{Energy landscape of a two-terminal conductor with a central scatterer. (a) Dc biased conductor: electrons within the bias window can be transmitted through the scatterer. 
(b)-(c) Examples of possible processes occurring in the presence of a time-dependent driving (shaded areas indicate the amplitude of the ac oscillations). In (b) an electron with energy in the bias window gets promoted above the chemical potential of the left reservoir by interacting with the ac-part of the driving. The corresponding vacancy does not contribute to transport as it cannot be filled by an electron tunneling from the right reservoir (i.e. it is reflected with probability 1). In contrast, in 
(c) both the electron as well as the remaining hole can be transmitted through the scatterer.}
\label{fig:sketch}
\end{figure}

Since electrons obey fermionic statistics, charge transfer across the sample can only occur if the incoming state is filled and, at the same time, the outgoing state is empty. A dc voltage bias applied to the conductor opens an energy window (typically named {\em bias window}) where both conditions are fulfilled and transport is possible, Fig.~\ref{fig:sketch}(a). 
An ac potential applied on top of the dc bias perturbs the local equilibrium of the lead it is applied to, creating electron-hole pair excitations.  Whether both the electron and the hole of a pair contribute to charge and energy transport depends on their energy 
with respect to the bias window, as illustrated schematically in some examples in Fig.~\ref{fig:sketch}(b)-(c). 

While electron-hole pairs will not contribute to the average charge current, it however has an important impact on the charge noise, limiting e.g. the precision of single-particle emitters. 
Moreover, the creation of an electron-hole pair is a correlated process and the manifestations of these correlations in physical observables are of particular interest. In Ref.~\onlinecite{Rychkov05}, Rychkov {\em et al.} developed an electron-hole description of the charge noise in linear response for coherent conductors that are subject to \textit{pure} ac voltages in the absence of a dc voltage.  This allows them to pinpoint contributions in the shot noise of a two-terminal conductor which are due to electron-hole correlations and coexist with electron-electron and hole-hole correlations. This result seemingly contradicts the interpretation given by Reydellet {\em et al.}\cite{Reydellet03} of their measurements of the shot noise in an ac driven point contact, which assumes that electrons and holes generate a partition noise independently from each other and  that the two contributions  add incoherently.  
Here, we generalise the approach of Rychkov {\em et al.}\cite{Rychkov05} to the case when the driving has an ac and a dc component of arbitrary magnitude, and show that the contradiction between Ref.~\onlinecite{Rychkov05} and  Ref.~\onlinecite{Reydellet03} is only apparent.   
Furthermore, we extend this description to energy transport and highlight the role of electron-hole correlations and independent contributions of electrons and holes to the energy noise.
 
In order to implement a description for electron- and hole-contributions to transport, we notice first that for the considered case of a periodically driven system, Eq.~\eqref{eq:currents}-\eqref{eq:noise-gen} can be re-written in energy space as\cite{Rychkov05} 
\begin{gather}
I_{\alpha}=\int 
dE \langle \hat{I}_{\alpha}(E) \rangle, \\ 
S_{\alpha\beta}=h \iint 
dE  dE' \langle \Delta \hat{I}_{\alpha}(E)\Delta \hat{I}_{\beta}(E') \rangle,
\end{gather}
with $\hat{I}_{\alpha}(E)=e/h\cdot \hat{i}_{\alpha}(E,E)$, and similarly for the average energy current $I_{\alpha}^{\E}$ and the energy noise $S_{\alpha \beta}^{\E}$,  with $\hat{I}_{\alpha}(E)$ simply replaced by $\hat{I}_{\alpha}^{\E}(E)=E/h\cdot \hat{ i}_{\alpha}(E,E)$. 
The energy-resolved current operators $\hat{I}_{\alpha}(E)$ and $\hat{I}_{\alpha}^{\E}(E)$   can be expressed as the sum of electron and hole  contributions
\begin{equation}\label{eq:eh-contrib}
\hat{I}_{\alpha}(E)=\sum_{i=\rm e,h}\hat{I}_{\alpha}^{ (i)}(E), \qquad \hat{I}_{\alpha}^{\E}(E)=\sum_{i=\rm e,h}\hat{I}_{\alpha}^{\E (i)}(E), 
\end{equation}
with
$\hat{I}_{\alpha}^{(i)}(E)=\hat{I}_{\alpha}(E)\theta_{i}(E)$, and $\hat{I}_{\alpha}^{\E (i)}(E)=\hat{I}_{\alpha}^{\E}(E)\theta_{i}(E) $.
Here, we introduced $\theta_{\rm e}(E)=\theta(E)$ and $\theta_{\rm h}(E)=\theta(-E)$, where $\theta(x)$ is the Heaviside step function. In other words we call electron-like (e-like) current the one carried by excitations with energy above the electrochemical potential $\mu$ of the right reservoir and hole-like (h-like) current the one carried by excitations with energy below $\mu$. This definition generalises the one given by Rychkov {\em et al.}\cite{Rychkov05} and it is motivated by the fact that an unoccupied electronic state created by the time-dependent driving in the bias window will never be able to participate in the transport as a hole because of Pauli exclusion principle, see Fig. \ref{fig:sketch}(b).  

For the average currents through the two-terminal conductor, the division in electron- and hole-like contributions, Eq.~\eqref{eq:eh-contrib}, leads straightforwardly to  $I^{\E}=I^{\E\mathrm{(e)}}+I^{\E\mathrm{(h)}}$, with ($i=\rm e,h$)
\begin{equation}
\begin{split}\label{eq:IE}
I^{\mathcal{E}(i)}=&\sum_{n} \int  \frac{dE}h E\, D_{n}(E)\left[  \tilde{f}_{\rm L}(E)-f_{\rm R}(E)\right]\theta_i(E),
\end{split}
\end{equation}
and $I^{\mathcal{E}(i)}\equiv I^{\mathcal{E}(i)}_{\rm R}=-I^{\mathcal{E}(i)}_{\rm L}$ because of the unitarity of the scattering matrix. 
The function $\tilde{f}_{\rm L}(E)=\sum_{\ell=-\infty}^{\infty} \left|c_{{\rm L}\ell}\right|^2 f_{\rm L}(E_{-\ell})$ is an effective non-equilibrium distribution function induced by the ac driving in the left reservoir, and it represents
the fact that in the presence of an ac-driving not only states with energy in the bias window contribute to transport, but also all their sidebands with energies
differing from $E$ by an integer multiple of $\hbar\Omega$. 
The coefficients $D_{n}(E)$ are eigenvalues of the matrix ${\bf s}_{\rm RL}^{\dag}(E){\bf s}_{\rm RL}^{}(E)$ that describes the transmission properties of the scatterer.  Apart from the non-equilibrium distribution function $\tilde{f}_{\rm L}$ replacing the equilibrium one $f_{\rm L}$, Eq.~\eqref{eq:IE} is formally identical to the expression for the energy current in a stationary conductor.\cite{Butcher90,vanHouten92,Guttman96} However, while in the stationary case at zero temperature there are either only e-like or only h-like excitations participating in the transport (depending on the sign of the bias voltage), in the ac driven case in general both kinds of excitations give a non vanishing contribution  $I^{\mathcal{E}(i)}_{\rm R}\Big|_{T=0}=\sum_{n} \sum_{\ell} \left|c_{{\rm L}\ell}\right|^2 \theta_i\big(\ell+\tfrac{e\bar{V}}{\hbar\Omega}\big) \int_0^{\ell\hbar\Omega+e\bar{V}} \!\!\! \tfrac{dE}{h} E\, D_{n}(E)$.
One remarkable exception is the case where $V_{\rm L}(t)$ has the form of
a series of Lorentzian pulses of quantized area ($\int eV(t)dt/h= N$, $N\in \mathbb{Z}$): in this case only the e-like or the h-like part of the current is non zero, depending on the polarity of the pulses.\cite{Keeling06,Dubois13nat,Dubois13}
The results and the discussion 
for the charge current are completely analogous, with the factor $E$ in the integrand of Eq.~\eqref{eq:IE} simply replaced by the electron charge $e$.

More insightful is the decomposition of the charge and the energy \textit{noise} into contributions that account for the correlations between the same or different types of excitations, e.g. $S^{\E}_{\alpha \beta}=\sum_{ij=\rm e,h}S^{\E (ij)}_{\alpha \beta}$ with
\begin{equation}\nonumber
S^{\E (ij)}_{\alpha \beta}=h\iint dEdE' \langle \Delta \hat{I}_{\alpha}^{\E (i)}(E)\Delta \hat{I}_{\beta}^{\E (j)}(E') \rangle,
\end{equation}
and similarly for the charge noise. A nonzero value for $S^{\E \rm(eh)}_{\alpha \beta}$ or $S^{\rm (eh)}_{\alpha \beta}$ is an unambiguous signature of the existence of intrinsic correlations between electron and hole excitations.\cite{Rychkov05}  

To simplify the notation, we will in the following use the fact that $S^{\mathcal{E}(ij)}_{\mathrm{RR}}=S^{\mathcal{E}(ij)}_{\mathrm{LL}}=-S^{\mathcal{E}(ij)}_{\mathrm{LR}}=-S^{\mathcal{E}(ij)}_{\mathrm{RL}}$ as a consequence of the unitarity of the scattering matrix  (and equivalently for the charge noise~\cite{Blanter00}), and always refer to the auto-correlators in the right reservoir $S^{\mathcal{E}(ij)}\equiv S^{\mathcal{E}(ij)}_{\mathrm{RR}}$ and $S^{(ij)}\equiv S^{(ij)}_{\mathrm{RR}}$. Interestingly,  in each of these quantities  
we identify two contributions with distinct character, i.e. $S^{\E (ij)}=S^{\E (ij)}_{\rm tr}+S^{\E (ij)}_{\rm int}$, with
\begin{subequations} \label{eq:SEij}
\begin{widetext}
\begin{gather}
S^{\mathcal{E}(ij)}_{\mathrm{tr}} = \frac1h  \sum_n \int dE\, E^2\,D_n(E)[1-D_{n}(E)] \left\{ \tilde{f}_\mathrm{L}(E)\left[1-f_\mathrm{R}(E)\right]+f_\mathrm{R}(E)\big[1-\tilde{f}_\mathrm{L}(E)\big]\right\} \theta_{i}(E)\theta_{j}(E),\\
S^{\mathcal{E}(ij)}_{\mathrm{int}}=\frac{1}{h}\sum_n\sum_{\alpha={\rm L,R}}\sum_{k,\ell,q=-\infty}^{\infty}c_{\alpha\ell}^{*}c_{\alpha(\ell+q)}c_{\alpha(k+q)}^{*}c_{\alpha k}\int dE\, EE_q D_{n}(E)D_{n}(E_{q})f_\alpha(E_{-\ell})\left[1-f_\alpha(E_{-k})\right]\theta_{i}(E)\theta_{j}(E_q).
\end{gather}
\end{widetext}
\end{subequations}
Similarly, for the charge noise we have $S^{(ij)}=S^{(ij)}_{\mathrm{tr}}+S^{(ij)}_{\mathrm{int}}$, where the expressions for $S^{(ij)}_{\mathrm{tr}} $and $S^{(ij)}_{\mathrm{int}}$ are formally analogous to Eq.~(\ref{eq:SEij}), but with $e^2$  replacing the factors $E^2$ and $E\, E_q$ in the integrands.

The two terms in Eq.~\eqref{eq:SEij} have different physical origins.\footnote{The distinction of transport-like contributions and interference terms enabled by the periodic driving, was not addressed in the linear-response regime considered in Ref.~\onlinecite{Rychkov05}.
} The first one stems  from correlations due to particle exchange between the two different reservoirs and we will therefore refer to it as \textit{transport part} of the noise. This is reflected in the fact that $S_{\rm tr}^{\E(ij)}$ depends on the occupation of both reservoirs, where the relevant energy distribution for the left one is  the non-equilibrium distribution $\tilde{f}_{\rm L}(E)$. 
Importantly, $S_{\rm tr}^{\E(ij)}$ is non zero only if one considers correlations between the same kind of excitations, i.e. $S_{\rm tr}^{\E\rm (eh)}=S_{\rm tr}^{\E\rm (he)}=0$. 
\begin{figure}[b]
\centerline{\psfig{figure=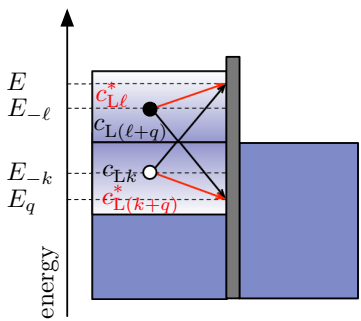,width=0.6\columnwidth}}
\caption{Sketch of a possible inelastic two-particle scattering event induced by the time-dependent driving applied to the left lead. The two reservoir states at energies $E_{-\ell}$ and $E_{-k}$ are connected to the incoming states at the scatterer at energies $E$ and $E_{q}$ by inelastic process via two possible pairs of energy-paths.}
\label{fig:interference}
\end{figure}

The correlations between electron- and hole-like excitations are therefore uniquely encoded  
 in the second term of Eq.~\eqref{eq:SEij}, $S_{\rm int}^{\E(ij)}$. This term originates ultimately from correlations due to the exchange of particles between states with different energies in the same reservoir.  Without periodic driving, only elastic exchange processes (i.e. thermal fluctuations)  contribute to the  
 noise, since $c_{{\rm L} k}=\delta_{k0}$ if $V_{\rm ac}=0$. In contrast, in the presence of a periodic driving, correlations between states with different energies  manifest themselves  in $S_{\rm int}^{\E(ij)}$ (and $S_{\rm int}^{(ij)}$), since in this case each state with a given energy impinging on the scatterer corresponds to a superposition of states with different energies propagating from the reservoirs (see Appendix).  Importantly, how much the correlations between the states with energies  $E_{-\ell}$ and $E_{-k}$ in reservoir L influence the charge and the energy noise depends on the interference between the different ``paths'' in energy space in which these states can contribute to states with energies $E$ and $E_{q}$ arriving at the scatterer, see Fig.~\ref{fig:interference}. For this reason, we name $S_{\rm int}^{\E(ij)}$ the {\em interference part} of the noise.

\section{Interference VS transport contributions to charge and energy noise}\label{sec:driving}

In order to make analytic progress and to compare with the existing literature on shot noise in ac-driven conductors,\cite{Blanter00,Rychkov05,Vanevic12} from now on we assume the central scatterer to have an energy-independent transmission. This assumption is very well suited e.g. for the experiment on noise in an ac driven two-terminal conductor with a tunnel barrier treated in Ref.~\onlinecite{Gabelli13}.   Moreover, to avoid overly cumbersome formulas, we restrict ourselves to the case of a spinless single-channel conductor.\footnote{The results can be easily generalised to account for spin degeneracy and several transmission channels by simply replacing $D(1-D)\to 2 \sum_nD_n(1-D_n)$ and $D^2\to 2 \sum_nD_n^2$.} Within these assumptions, we will here discuss in detail the interference and the transport part of the charge and energy noise  introduced in Sec.~\ref{sec:ehpicture}, and their physical interpretation.  

When referring to specific driving signals, we use as examples a simple harmonic driving, a Lorentzian-shaped signal, leading to quantized charge emission, as well as a bi-harmonic driving signal, which has recently been shown to be an intriguing experimental way of approaching the case of Lorentzian driving via the superposition of harmonic signals.
\subsection{Interference part}\label{subsec_interference}
\begin{figure}[t,b]
\centerline{\psfig{figure=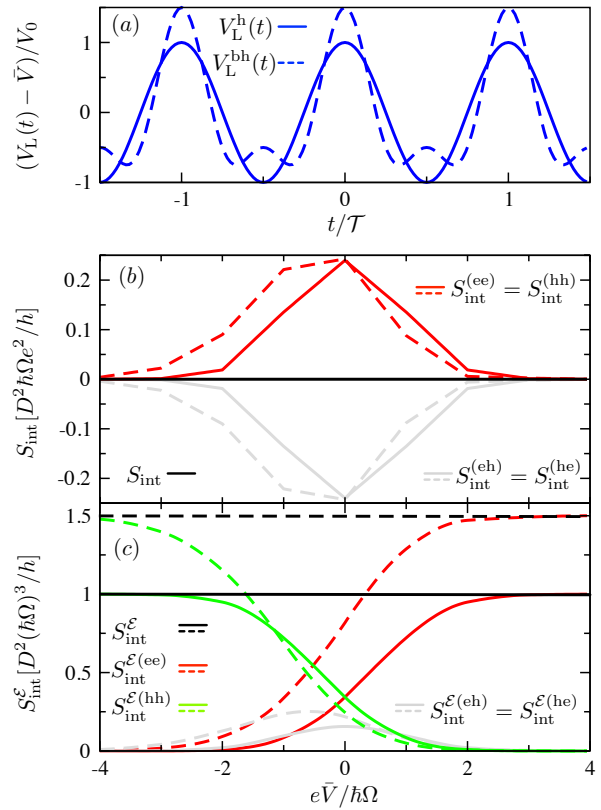,width=0.95\columnwidth}}
\caption{ Plots of the interference contributions to the charge and energy noise as a function of the dc offset of the driving.   (a) Line-shape of the applied voltages: harmonic driving, $V^\text{h}_{\rm L}(t)=\bar{V}+V_{0}\cos(\Omega t)$, and bi-harmonic driving of the form $V^\text{bh}_{\rm L}(t)=\bar{V}+V_{0}\cos(\Omega t)+\frac{V_{0}}2\cos(2\Omega t)$.
(b) Interference contributions to the charge noise  $S_{\rm int}^{(ii)}$ and $S_{\rm int}^{(i\neq j)}$ as well as their sum {$S_\text{int}$}.   (c) Interference contributions to the energy noise $S_{\rm int}^{\E (ii)}$ and $S_{\rm int}^{\E (i\neq j)}$, as well as  the total contribution of the interference terms {$S_{\rm int}^{\E}$}. In all panels $k_{\rm B}T=0$ and $eV_{0}=2\hbar\Omega$. Full lines correspond to the case of harmonic driving while dashed lines represent the case of bi-harmonic driving.}
\label{fig:Sint}
\end{figure}

We start by considering the interference part of the noise,  $S^{(ij)}_{\rm int}$ and $S^{\E(ij)}_{\rm int}$. 
As already noticed  in Sec.~\ref{sec:ehpicture}, these interference terms are the only ones that contribute to 
the mixed correlators between electron and hole currents, i.e. $S^{\rm (eh)}=S^{\rm (eh)}_{\rm int}$  and $S^{\E\rm (eh)}=S^{\E\rm (eh)}_{\rm int}$. A non-zero value for $S^{\rm (eh)}$  and $S^{\E\rm (eh)}$ is a clear fingerprint of the existence of intrinsic correlations between the electron and hole excitations created by the ac driving, as discussed in detail by Rychkov {\em et al.} in the case of pure ac driving. \cite{Rychkov05} These correlations however quickly decay for increasing amplitude of the dc component $\bar{V}$ of the driving, as  illustrated in Fig.~\ref{fig:Sint}(b-c), where we plot $S^{(ij)}_{\rm int}$ and $S^{\E (ij)}_{\rm int}$  for two different driving fields, $V^\text{h}_{\rm L}(t)=\bar{V}+V_{0}\cos(\Omega t)$ (harmonic  driving, full lines) and $V^\text{bh}_{\rm L}(t)=\bar{V}+V_{0}\cos(\Omega t)+\frac{V_{0}}2\cos(2\Omega t)$ (bi-harmonic  driving, dashed lines), as a function of the dc component of the bias at zero temperature.

The interference terms, Eq.~(\ref{eq:SEij}b), do not only contribute to the correlators between electrons and holes, but  also to those between the same type of particles, i.e. to $S^{(ii)}$ and $S^{\E(ii)}$. 
Interestingly, independently of the symmetry of the driving signal and of the amplitude of the dc component,  the interference contribution to the electron-electron and hole-hole correlators in the charge noise are equal and opposite in sign to the correlators between different types of excitations, i.e. 
$S^{\rm(ee)}_{\rm int}=S^{\rm(hh)}_{\rm int}=-S^{\rm(eh)}_{\rm int}=-S^{\rm(he)}_{\rm int}$, see Fig.~\ref{fig:Sint}(b). The total contribution to the charge noise due to interference terms  at zero temperature is therefore identically zero. More in general, at finite temperature one can show that the total contribution of the interference terms to the charge noise reduces to temperature fluctuations
\begin{equation}
\begin{split}\label{eq:Sint-tot}
S_{\rm int}=\sum_{ij}S_{\rm int }^{(ij)}=2\frac{e^{2}}{h}D^{2}k_{\rm B}T,
\end{split}
\end{equation}
since all paths in energy space that correspond to inelastic processes interfere destructively when they contribute 
to a physical observable with the same weight ($e^{2}$, for the case of the charge noise).

The situation is different for what concerns the interference contribution to \textit{energy noise}. In this case we obtain 
\begin{align}\label{eq:absorb}
S^{\E}_{\rm int}=\sum_{ij}S_{\rm int }^{\E(ij)} &=\frac{2\pi^{2}}{3h}D^{2}\left(k_{\rm B}T\right)^3\\
&+\frac{D^{2}}{2h}\sum_{k=-\infty}^{+\infty}\left|e\, v_{{\rm L}k}\right|^{2}k\hbar\Omega\coth\left(\frac{k\hbar\Omega}{2k_{\rm B}T}\right),\nonumber
\end{align}
with  $v_{{\rm L}k}=\int_{0}^{\T}\frac{dt}\T V_{\rm L}(t)e^{ik\Omega t}$.  The first term is similar to Eq.~\eqref{eq:Sint-tot}
 and it is purely due to thermal fluctuations. The second one instead results from the probabilistic absorption of energy from the ac driving field.
Following the line of known results for the finite-frequency charge noise, see e.g. Ref.~\onlinecite{Gavish04PhD}, the factor 
$\frac{D^{2}}{h}k\hbar\Omega\coth\left(\frac{k\hbar\Omega}{2k_{\rm B}T}\right)$ can in fact be interpreted as the characteristic rate at which electrons exchange the energy $k\hbar\Omega$ with the ac fields by fluctuating between states in the {same} reservoir.
\footnote{This can be best understood by comparing the form of this term with the finite-frequency {\em charge} noise $S(\omega)$ of a dc-biased system. It is well understood that $S(\omega)$ is directly related to the characteristic rate at which the system exchanges energy with the environment,  $S(\omega)\propto \Gamma^{\rm emis}_{\rm tot}(\omega)+\Gamma^{\rm abs}_{\rm tot}(\omega)$, where $\Gamma^{\rm emis (abs)}_{\rm tot}(\omega)$ is the total probability per unit time that the system emits (absorbs) the energy $\hbar\omega$ from the environment (see e.g. Ref.\onlinecite{Gavish04PhD} and references therein).  Here, the term $\frac{e^2}{h}D^{2}k\hbar\Omega\coth\left(k\hbar\Omega/2k_{\rm B}T\right)$ is the part of the interference contribution to the zero-frequency noise $S$ that accounts  for the energy exchange of $k\hbar\Omega$ due to fluctuations between states in the {\em same} reservoir. In our case the role of the environment is played by the classical ac driving field.}  
The factor $\left|e\, v_{{\rm L}k}\right|^{2}/2$
 is equal to the variance of the energy of a classical, charged particle in the oscillating potential $v_{{\rm L} k}\cos(k\Omega t)$, i.e. the $k$th mode of the periodic driving potential. 
 
 The difference between the interference contributions to the charge and energy noise is obvious also when considering the {different} contributions $S^{(ij)}_{\rm int}$ and $S^{\E (ij)}_{\rm int}$, as shown in Fig.~\ref{fig:Sint}. {While for} the interference part to the charge noise at zero temperature we had $S^{\rm (ee)}_{\rm int}=S^{\rm (hh)}_{\rm int}$  for all values of $\bar{V}$;
for the energy noise we find in general $S_{\rm int}^{\E \mathrm{(ee)}}\gtrless S_{\rm int}^{\E \mathrm{(hh)}}$, depending on the polarity of the dc component $\bar{V}$ as well as on the shape of the ac-part of the driving, i.e. depending on whether $\max{V_{\rm ac}(t)}\gtrless |\min{V_{\rm ac}}(t)|$. Around $\bar{V}=0$, the mixed correlator $S^{\E \rm (eh)}_{\rm int}$ has roughly the same order of magnitude as $S_{\rm int}^{\E \mathrm{(ee)}}$ and $S_{\rm int}^{\E \mathrm{(hh)}}$, but it decays quickly as soon as the dc part of the bias is increased, indicating a suppression of the correlations between electron and hole excitations for large $\bar{V}$. For large values of $\bar{V}$, $S^{\E }_{\rm int}$ consists essentially only of the contribution of one type of excitations, namely electrons for $e\bar{V}>0$ and holes for $e\bar{V}<0$.

In Fig.~\ref{fig:Sint}, we choose a biharmonic driving signal where the first and second harmonic oscillate in phase.  Bi-harmonic driving signals of the shape $V^\text{bh}_{\rm L}(t)=\bar{V}+V_{1}\cos(\Omega t)+V_{2}\cos(2\Omega t+\varphi)$ were used in the experiment of Ref.\onlinecite{Gabelli13} to minimize the charge noise by tuning the phase $\varphi$. It is an important property of the interference contribution of the energy noise that it is fully insensitive to the phase difference $\varphi$ between the different harmonics, see Eq.~(\ref{eq:absorb}). It is furthermore insensitive to the dc offset and only exposes the amplitudes of the different harmonics, $V_1$ and $V_2$, and their frequencies. 

Importantly, for the case of a perfectly transmitting conductor ($D=1$), the second term of Eq.~(\ref{eq:absorb}) is the only contribution to the energy noise that survives  at zero temperature. It is hence expected to be well observable for scatterers with a high transmission, where the transport part is suppressed.

\subsection{Transport part}\label{subsec_transport}
We now consider the \textit{transport contributions} to the charge and energy noise, $S_{\rm tr}=\sum_{i}S^{(ii)}_{\mathrm{tr}}$ and $S^{\E}_{\rm tr}=\sum_{i}S^{\mathcal{E}(ii)}_{\mathrm{tr}}$, which, for the case of a conductor with energy independent transmission,  are given by
\begin{align} 
S_{\mathrm{tr}} &= \frac{e^2}{h}D(1-D) \nonumber \\
&\times \sum_{\ell=-\infty}^{+\infty}|c_{\rm L \ell}|^{2}(\ell\hbar\Omega+e\bar{V})\coth\left(\frac{\ell\hbar\Omega+e\bar{V}}{2k_{\rm B}T}\right), \label{eq:StrT}\\
S^{\mathcal{E}}_{\mathrm{tr}}&= \frac{D(1-D)}{3h} \sum_{\ell=-\infty}^{+\infty}|c_{\rm L \ell}|^{2}\coth\left(\frac{\ell\hbar\Omega+e\bar{V}}{2k_{\rm B}T}\right)\nonumber \\
&\times \left[(\ell\hbar\Omega+e\bar{V})^3+(\ell\hbar\Omega+e\bar{V})(\pi k_{\rm B}T)^2\right].\label{eq:SEtrT}
\end{align}

As remarked in Sec.~\ref{sec:ehpicture}, the transport part of the noise sees no signatures of the correlations between electrons and hole excitations, which means in turn that the two types of excitations contribute {\em independently} to the transport part of the noise. For this reason, we  can interpret Eq.~\eqref{eq:StrT}-\eqref{eq:SEtrT} in a rather classical, ``particle-like'' picture and associate the factor $D(1-D)$ with the partition noise of the scatterer caused by the random transmission/reflection of charge carriers.  

It is important to notice that at zero-temperature $S_{\mathrm{tr}}$ represents the only non vanishing contribution to the charge noise. It follows that,  although the electron and hole excitations created by the ac driving are in general {\em not} independent, as pointed out by Rychkov {\em et al.}\cite{Rychkov05} and discussed in Sec.~\ref{subsec_interference}, nevertheless the {\em total} charge noise in the two-terminal conductor at zero temperature can be written 
as the incoherent sum of a contribution due to electron-like excitations and one due to hole-like excitations, as first suggested by Reydellet {\em et al.}\cite{Reydellet03}  
\begin{equation}\label{eq:SiiT0}
S(k_{\rm B}T=0)=\sum_{i=\rm e,h}\frac{e^{2}\Omega}{2\pi}D(1-D)N_{i}\ ,
\end{equation}
where 
\begin{equation}
N_{i}=\sigma_{i}\sum_{\ell=-\infty}^{+\infty}|c_{\rm L \ell}|^{2}(\ell+e\bar{V}/\hbar\Omega)\theta_{i}(\ell \hbar\Omega+e\bar{V}),
\end{equation}
is the average number of electrons or holes that impinge on the scatterer during one period ($\sigma_{e/h}=\pm$). 
If the driving $V_{\rm L}(t)$ has a non-zero dc component, the number of electron or hole excitations attempting to cross the barrier per period will be different, with $N_{\rm e}\gtrless N_{\rm h}$ depending on $e\bar{V}\gtrless0$.

 It is interesting to relate Eq.~\eqref{eq:SiiT0} to the picture of charge transport in ac driven conductors drawn by Vanevi\'c  and coworkers.\cite{Vanevic07,Vanevic08,Vanevic12}  By investigating the full  counting statistics of an energy-independent scatterer, they classified the elementary charge transport processes occurring in the presence of a time-dependent driving in ``unidirectional'' and ``bidirectional events''.\cite{Vanevic07,Vanevic08} The first  ones have a single-particle-like character and their counting statistics is equivalent to the one of electrons (or holes) transmitted through a barrier with a static voltage drop, while the second ones describe neutral excitations with pair-like character. According to the definition of electron and hole excitations given in Sec.~\ref{sec:model}, if we assume for example $e\bar{V}>0$, we have that $N_{e}$ accounts both for the electrons injected by the dc bias and for those that are part of electron-hole pairs created by the ac driving. In contrast, since for $e\bar{V}>0$ holes can be excited only by the ac-part of the driving, $N_{\rm h}$ corresponds exactly  to the number of those electron-hole pairs where the electron- and the hole-like excitation have the same probability $D$ to be transmitted through the barrier.  The number of electron excitation that participate in the transport {\em without} 
 a hole-counterpart is then  $(N_{e}-N_{h})=e\bar{V}/\hbar\Omega$. Rewriting Eq.~\eqref{eq:SiiT0} as
\begin{equation}\label{eq:StrT0}
S(k_{\rm B}T=0)=\frac{e^{2}\Omega}{2\pi}D(1-D)[(N_{\rm e}-N_{\rm h})+2N_{\rm h}],
\end{equation} 
we can associate the first term, which correspond to the noise due to the ``unpaired'' electron-like excitations (i.e. those that are injected by the dc bias and do not interact with the ac driving, as well as those which were originally in the bias window and got promoted above the Fermi level of the left reservoir by absorbing a certain amount of Floquet quanta), to the unidirectional events of  Vanevi\'c  {\em et al.}\cite{Vanevic07,Vanevic08,Vanevic12} The second one is associated to bidirectional events, i.e. to the excess noise due to neutral pair-excitations created by the ac-part of the driving. The factor 2 accounts for the fact that the electron and the hole of a pair contribute equally to the charge noise. Note that it is a consequence of considering a conductor with energy-independent transmission that the charge noise due to ``unpaired'' excitations (unidirectional events) is equal to the one that would occur in the presence of a dc bias only. In this case in fact, the two processes represented schematically in Fig,~\ref{fig:sketch}(a)-(b) are completely equivalent from the point of view of charge transport.

\begin{figure}[t,h]
\centerline{\psfig{figure=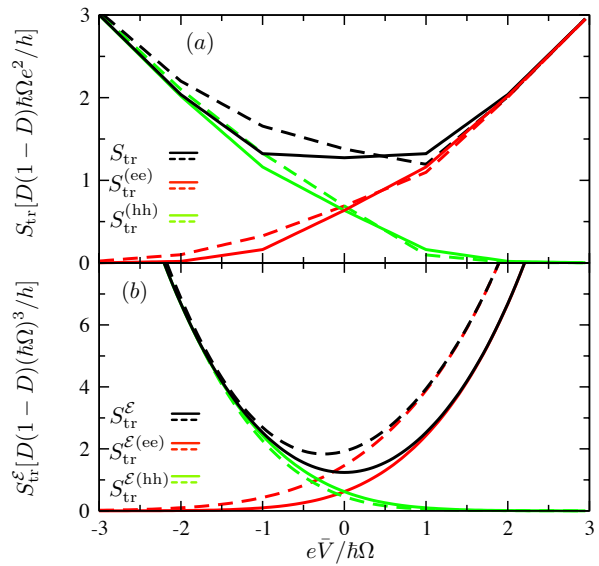,width=0.95\columnwidth}}
\caption{Plots of the transport contributions to the charge and energy noise as a function of the dc offset of the driving.  In both panels full lines correspond to the case of harmonic driving $V_{\rm L}^{\rm h}(t)=\bar{V}+V_{0}\cos(\Omega t)$ and dashed lines to bi-harmonic driving of the form $V^{\rm bh}_{\rm L}(t)=\bar{V}+V_{0}\cos(\Omega t)+\frac{V_{0}}2\cos(2\Omega t)$, with $eV_{0}=2\hbar\Omega$. 
(a) Transport contributions to the charge noise  $S_{\rm tr}^{(ii)}$ and $S_{\rm tr}=\sum_{i}S_{\rm tr}^{(ii)}$ . (b) Transport contributions to the energy noise $S_{\rm tr}^{\E (ii)}$ and $S_{\rm tr}^{\E}=\sum_{i}S_{\rm int}^{\E (ii)}$. In all panels $k_{\rm B}T=0$. }
\label{fig:Str}
\end{figure}

The approach of Vanevi\'c {\em et al.}\cite{Vanevic07,Vanevic08} has been successfully applied to interpret noise measurements in ac driven conductors,~\cite{Vanevic12} showing in particular how the reduction of the shot noise reported by Gabelli and Reulet~\cite{Gabelli13} in a tunnel junction driven by a bi-harmonic signal is related to a suppression of the probability of exciting pair-like neutral excitations.  
The suppression of the charge noise obtained with a bi-harmonic modulation is illustrated in Fig.~\ref{fig:Str}a. Here we plot the transport contribution to the charge correlators  $S_{\rm tr}^{(ii)}$ at zero temperature, for the same kind of harmonic and bi-harmonic driving considered in Fig.~\ref{fig:Sint}. For the case of bi-harmonic driving  the minimum of the total charge noise $S=S_{\rm tr}$ is situated at finite $\bar{V}$ and, most interestingly, this minimum can be smaller than the one obtained in the presence of a simple harmonic driving. If one looks at the individual contributions due to (ee) and (hh)-correlations, one sees that the minimum of the noise corresponds to a suppression of both $S^{\rm(ee)}_{\rm tr}$ and $S^{\rm(hh)}_{\rm tr}$ with respect to the case of harmonic driving, leaving their \textit{difference} unchanged and thereby indicating that the reduction of the charge noise is indeed related to a smaller number of electron-hole excitations participating to transport, as discussed in detail in Refs.~\onlinecite{Gabelli13,Vanevic12}. The reduction of the charge noise for signals designed to approach the Leviton case by superposing several harmonics has been shown in Ref.~\onlinecite{Dubois13nat}. 

 The suppression of the charge noise obtained with bi-harmonic driving has in general no counterpart in the energy noise $S^{\E}_{\rm tr }$, see Fig.~\ref{fig:Str}(b).  This is because, while the charge noise depends only on the total \textit{number} of excitations created by the driving (in the case of a scatterer with energy-independent transmission), the energy noise is sensitive to their \textit{energy distribution}. The latter depends sensitively on the shape of the ac driving and on its dc component $\bar{V}$.  For example, the bi-harmonic signal $V_{\rm L}^{\rm bh}(t)$ considered for the curves of  Fig.~\ref{fig:Str} has larger excursions for positive than for negative values when taking $\bar{V}=0$, see Fig.~\ref{fig:Sint}(a). As a consequence, around $\bar{V}=0$ it creates electronic excitations with a larger spread in energy than the corresponding holes, resulting in $S_{\rm tr}^{\E{\rm (ee)}}>S_{\rm tr}^{\E{\rm (hh)}}$. 
Moreover, increasing the dc component of the driving enlarges both the number of one type of excitations participating in transport (e.g. $N_{\rm e}$ for $e\bar{V}>0$) as well as the energy range that they span, while it suppresses the contribution of the other type of carriers. 

Therefore a minimum of the transport part of the energy noise occurs when a trade-off between the amount of excited particles and their spread in energy is reached. In other words, in order to minimize the transport part of the energy noise, it can be favourable to even \textit{increase} the amount of excited particles, if only the total energy spread is reduced at the same time. This is in contrast to the charge  noise, where a minimum is reached when the {\em number} of excited particles is minimized. Hence, when comparing Figs.~\ref{fig:Str}(a) and (b) the minima of $S_\mathrm{tr}$ and $S^\E_\mathrm{tr}$ are found for average bias voltages $\bar{V}$ of opposite sign.

For sufficiently large values of $\bar{V}$ we can consider essentially only one type of excitation, as can be seen from Fig.~\ref{fig:Str}(b). Then,    
 $S^{\E}_{\rm tr}$ at zero temperature takes the simple form
\begin{equation} \label{eq:SEapprox}
S_{\rm tr}^{\E}\approx \frac{D(1-D)}{h}\frac{\overline{|e V_{\rm L}(t)|^{3}}}3,
\end{equation}
which resembles closely the form of the  
energy noise in the case of a pure dc bias applied to the conductor 
\begin{equation}\label{eq:SEdc}
S^\mathcal{E}(k_\mathrm{B}T=0, V_{\rm ac}(t)=0)  = \frac{D(1-D)}{h}\frac{\left|e\bar{V}\right|^3}3.
\end{equation}
In other words, for large dc bias,  $S_{\rm tr}^{\E}$ is equal to the time average of the energy noise that one would obtain in a series of ``frozen frames'' with a static bias potential, indicating that in this case the transport part of the energy noise can be interpreted as the contribution of particle excitations of only one kind that follow \textit{instantaneously} the modulation of the driving potential $V_{\rm L}(t)$. 

This simple picture breaks down as soon as $\bar{V}$ is small enough so that $V_\mathrm{L}(t)$ changes sign during the driving period.
Then both  $S^{\E \mathrm{(hh)}}_{\rm tr}$  and $S^{\E \mathrm{(ee)}}_{\rm tr}$ are non vanishing, and $S^{\E}_{\rm tr}$ deviates from Eq.~\eqref{eq:SEapprox}.  This is different from what one observes for the energy current, $I^\mathcal{E}=D \overline{(eV(t))^2}/2h$, which can always be seen as the time-average of the energy current from a series of ``frozen frames'' with a static bias potential. The reason is that while the current is a single-time operator, the noise depends  on the correlations between the current operator at two different times and, if the sign of $V_{\rm L}(t)$ changes over time, there is no simple mapping between the energy space, in which we define electron-like and hole-like excitations, and the time-space, in which one defines the ``frozen frames''.

\begin{figure}[b]
\centerline{\psfig{figure=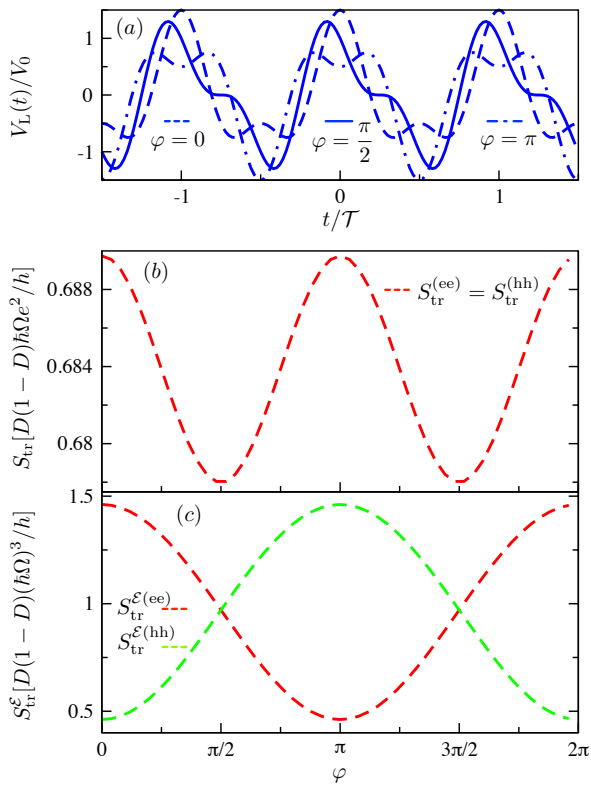,width=0.95\columnwidth}}
\caption{ (a) Shape of the ac excitation  $V_{\rm L}(t)=V_{0}\cos(\Omega t)+\frac{V_{0}}2\cos(2\Omega t+\varphi)$ for different phase shifts: $\varphi = 0$ (dashed), $\varphi = \pi/2$ (full), $\varphi = \pi$ (dash-dotted). (b)-(c)
Plots of the transport contributions to the charge and energy noise for the type of bi-harmonic driving introduced in (a), as a function of the phase $\varphi$ between the two harmonic components.   In all panels $eV_{0}=2\hbar\Omega$ and $k_{\rm B}T=0$. }
\label{fig:phase_dep}
\end{figure}

For small dc bias $\bar{V}\approx 0$,  the energy noise of the electron- and the hole-like excitations depends strongly on the shape of $V_{\rm ac}(t)$.  This can be seen for example in Fig.~\ref{fig:phase_dep}, where we plot $S^{(ii)}_{\rm tr}$ and $S^{\E (ii)}_{\rm tr}$  for the case of bi-harmonic driving  $V_{\rm L}(t)=V_{0}\cos(\Omega t)+\frac{V_{0}}2\cos(2\Omega t+\varphi)$ as a function of the relative phase $\varphi$ between the two harmonics. Changing $\varphi$ one obtains different waveforms for  $V_{\rm L}(t)$, see Fig.~\ref{fig:phase_dep}(a). 
We notice that while for the charge noise the transport contributions due to electron and hole excitations are equal for all $\varphi$, for the energy noise we have in general $S^{\E \rm (ee)}_{\rm tr}\neq S^{\E \rm (hh)}_{\rm tr}$. This is because, while a pure ac voltage $V_{\rm L}(t)=V_{\rm ac}(t)$ always creates the same number of electron and hole excitations, the energy distribution of the two types of excitations is equal only if $V_{\rm L}(t)$ is antisymmetric with respect to one of its nodes. For the kind of driving considered in Fig.~\ref{fig:phase_dep}, this occurs for $\varphi=(2N+1)\pi/2$, $N\in \mathbb{Z}$. For other values of $\varphi$,  $eV_{\rm L}(t) $ has larger excursions for either positive or negative values and it therefore creates distributions of either electrons or holes that have a larger spread in energy than the one of the other type of carriers. At $\bar{V}=0$ the total transport contribution to the energy noise $S_{\rm tr}^{\E}=\sum_{i=\rm e,h}S_{\rm tr}^{\E (ii)}$ is however independent of $\varphi$, see Fig.~\ref{fig:phase_dep}. At finite bias also the total transport part of the energy noise depends on $\varphi$. This is in stark contrast to the interference contribution to the energy noise, which is always independent of the phase dependence between the two superposed harmonics.

\subsection{Lorentzian pulses}
Based on the insights on the transport and interference parts of charge and energy noise, we finally address the case of a time-dependent driving with Lorentzian shape, 
\begin{equation}\label{eq:Lev}
V_\mathrm{L}(t) = \frac{V_0}{\pi}\sum_{q=-\infty}^\infty\sum_{r=1}^M\frac{\mathcal{T} W}{(t-(q+ d_r)\mathcal{T})^2+W^2}.
\end{equation}
This rather general notation describes a series of $M\in\mathbb{N}$ Lorentzian pulses emitted at fractions $0<d_r<1$ of the period, with a full width at half of the maximum value equal to $2W$. The single pulses have a quantized area if $eV_0=N\hbar\Omega$ with $N\in\mathbb{Z}$. Restricting ourselves to the case of zero temperature and assuming that the width of the pulses is much smaller than their distance, avoiding a possible overlap, it is known that a sequence of pulses with quantized area leads to the injection of an integer number of particles on the scatterer, without copropagating electron-hole pairs.~\cite{Keeling06} This leads to a suppression of the charge noise, which is reduced to the level of a purely dc-biased conductor, $S=S_\text{tr}=\frac{e^2}{h}D(1-D)|eV_0|$.~\cite{Ivanov97,Keeling06,Keeling08,Dubois13nat,Dubois13}  
However, both the interference and the transport part of the energy noise have considerable contributions of comparable magnitude, related to the specific shape of the applied voltage pulse. 

At zero temperature and for an integer number of particles injected on the scatterer by the driving potential given in Eq.~(\ref{eq:Lev}), the transport and interference part of the energy noise allow for a rather intuitive interpretation. The absence of electron-hole pairs yields a transport part of the energy noise, $S^\E_\mathrm{tr}=D(1-D)\overline{|e V_{\rm L}(t)|^{3}}/3h$, which indicates that the particles injected on the scatterer follow the potential modulation instantaneously, see the discussion of Eq.\eqref{eq:SEapprox}. Moreover, the interference part yields a contribution inversely proportional to the square of the width of the pulse, which is related to the intrinsic energy spread of the emitted particles. In particular we find $S^\E_\mathrm{int}(N=M=1)=D^2(\hbar/2W)^2/\mathcal{T}$, if one particle is injected per period.~\cite{Moskalets14}

Importantly, it is known that $L$ single particles injected by separate pulses are truly independent of each other.~\cite{Keeling06} Such a situation is represented by the pulses given in Eq.~(\ref{eq:Lev}) for $N=\pm1$ and arbitrary values of $M=L$. This contrasts with the situation where several particles are injected by the same pulse, namely $|N|=L$: while also in this case no electron-hole pairs contribute to transport, the emitted particles are however \textit{not} independent of each other.~\cite{Ivanov97} This fact is not visible from the charge current and charge noise: in fact $I(N=L,M=1)=I(N=1,M=L)=LI(N=1,M=1)$ and $S(N=L,M=1)=S(N=1,M=L)=LS(N=1,M=1)$. This is  different for the energy current and the total energy noise, where these two cases can indeed be distinguished. For a series of $L$ Lorentzian pulses carrying each one single particle ($N=\pm1$ and $M=L$), the energy current and the total energy noise are given by the sum of the contributions of the single particles carried by separate pulses.  We hence have $I^\E(M=L)=LI^\E(M=1)$ as well as $S^\E(M=L)=LS^\E(M=1)$, indicating the independence of the injected particles, also called ``Levitons".~\cite{Dubois13nat}  
However, when $L$ particles are emitted in the same pulse ($|N|=L$ and $M=1$) the energy current~\cite{Moskalets09} and energy noise of the particles do not simply add up any longer. This can be seen from $I^\E(N=L)=L^2I^\E(N=1)$ for the energy current and from $S^\E_\mathrm{int}(N=L)=L^2S^\E_\mathrm{int}(N=1)$ and $S^\E_\mathrm{tr}(N=L)=|L^3|S^\E_\mathrm{tr}(N=1)$ for the two contributions to the energy noise. The total energy noise is hence not  proportional to the energy noise of a single Leviton any longer.

\section{Heat and power fluctuations}\label{sec:power}

For a realistic measurement, it is the \textit{heat} current which is of relevance, rather than the \textit{energy} current. The heat current operator
\begin{eqnarray}\label{eq:heatoperator}
\hat{J}_\alpha(t) & = & \hat{I}^\mathcal{E}_\alpha(t)-V_\alpha(t)\hat{I}_\alpha(t)
\end{eqnarray}
accounts for the flow of energy in each reservoir that is in excess with respect to the local electrochemical potential and that must be dissipated in some nearby thermostat to keep the reservoirs in thermal equilibrium. Since neither charge nor energy can be accumulated in the conductor, the sum of the heat currents flowing in all the contacts must be equal to the total power injected by the voltage source into the system. At the operator level, this means for a two-terminal conductor
\begin{equation}\label{eq:poweroperator}
\hat{J}_\mathrm{L}(t)+\hat{J}_\mathrm{R}(t)= \hat{P}(t)
\end{equation}
with $\hat{P}(t)=-V(t)\hat{I}_\mathrm{L}(t)$ the operator for the power provided by the time-dependent voltage source.  From Eq.~\eqref{eq:poweroperator} it follows directly that $S_\mathrm{LL}^{J}+S_\mathrm{LR}^{J}+S_\mathrm{RL}^{J}+S_\mathrm{RR}^{J}=\langle(\Delta P)^2\rangle$, where 
\begin{eqnarray}
S_{\alpha\beta}^{J} &=& \frac{1}{2\mathcal{T}}\int_{0}^{\mathcal{T}}\!\! dt\int_{-\infty}^{\infty}\!\! d\tau\langle\{\Delta \hat{J}_{\alpha}(t),\Delta \hat{J}_{\beta}(t+\tau) \} \rangle ,
\end{eqnarray}
is the \textit{heat noise} and 
\begin{align}
&\langle (\Delta P)^{2}\rangle=\frac{1}{2\mathcal{T}}\int_{0}^{\mathcal{T}}dt\int_{-\infty}^{\infty}d\tau\langle\{\Delta \hat{P}(t),\Delta \hat{P}(t+\tau) \} \rangle \nonumber \\
&=\frac{1}{2\mathcal{T}}\int_{0}^{\mathcal{T}}dt\int_{-\infty}^{\infty}d\tau V(t)V(t+\tau)\langle\{\Delta \hat{I}(t),\Delta \hat{I}(t+\tau) \} .
\end{align}
is the  zero-frequency correlator of power fluctuations. The latter is an experimentally accessible quantity via measurements of the current correlator and the applied voltages.~\footnote{J. Gabelli, private communication.} The power fluctuations are directly related to the fluctuations in the work done by the time-dependent voltage applied across the sample during a long-time measurement.
The features in the energy noise discussed in Sec.~\ref{sec:driving} can be detected in the fluctuations of the power delivered by the voltage source. Explicitly, we find
\begin{align}
&\langle (\Delta P)^{2}\rangle =\frac{D^2}{h}  \sum_{k=-\infty}^{+\infty}|ev_{\rm L k}|^{2}k\hbar\Omega\coth\left(\frac{k\hbar\Omega}{2k_{\rm B}T}\right)\nonumber \\
&+\frac{D(1-D)}{h} \sum_{\ell=-\infty}^{+\infty}|c_{\rm L \ell}|^{2}(\ell\hbar\Omega+e\bar{V})^3\coth\left(\frac{\ell\hbar\Omega+e\bar{V}}{2k_{\rm B}T}\right)\label{eq:DeltaP2}
\end{align}
which at zero temperature reduces to
\begin{eqnarray}
\langle (\Delta P)^{2}\rangle=2S_{\rm int}^{\E}(k_{\rm B}T=0)+3S_{\rm tr}^{\E}(k_{\rm B}T=0)\ 
\end{eqnarray}
containing the interference and transport part of the energy noise discussed in this paper. Importantly, interference and transport part of the noise can be distinguished by their different dependence on the transmission of the barrier, which in systems containing quantum point contacts can be made tunable. For weakly transmitting barriers, the transport part is the main contribution, while for transparent barriers only the interference part matters.~\footnote{\red{Note that Eq.\eqref{eq:DeltaP2} corresponds to twice the heat noise $\mathcal{S}_{Q}$ defined in Eq.(8c) of Ref.~\onlinecite{Moskalets14}, i.e. $\langle (\Delta P)^{2}\rangle=2 \mathcal{S}_{Q}$. However, for the power fluctuations of a series of Lorentzian pulses with unit flux at zero temperature we obtain $\langle (\Delta P)^{2}\rangle =2 (\hbar/2W)^{2}D[1+2(1-D)]/\mathcal{T}$, in contrast with $\mathcal{S}_{Q}=(\hbar/2W)^{2}D/\mathcal{T}$  predicted in Ref.~\onlinecite{Moskalets14}. This discrepancy stems from a mistake in the calculations of Ref.~\onlinecite{Moskalets14} 
and contradicts the conclusion of Ref.~\onlinecite{Moskalets14} that  the excitations generated by a series of Lorentzian voltage pulses with quantized flux carry a heat noise that does not grow upon reflection.}}

\section{Conclusions}

We discussed the charge and energy transport properties of a two-terminal conductor driven by an arbitrary time-periodic bias voltage, focusing in particular on the role of electron- and hole-excitations in the charge and energy noise. The noise exhibits two contributions of different character, namely transport contributions, containing correlations between particles in different leads, and interference contributions, due to correlations between particles in the same lead. We could show that only the latter contains signatures of the correlations between electrons and holes excited by the time-dependent driving. For the charge noise the interference part vanishes at zero temperature, allowing for an interpretation of the total charge noise in terms of binomial statistics of independent charge carriers, only. The total charge noise can hence be expressed in terms of the numbers of electron and hole excitations.
  
The properties of the energy noise are more intricate, revealing quantum effects due to the superposition of different energy states in its interference part. The latter can be related to the variance of the energy distribution of  the excitations due to different modes of the ac-part of the driving potential. Furthermore, in contrast to the charge noise, the transport part of the energy noise is sensitive to the \textit{spread in energy} of the excited particles rather than to their \textit{number}. Therefore, a minimum in the charge noise, occurring when the number of electron-hole pair excitations is suppressed, does not go along with a minimum in the energy noise. The minimum in the energy noise is due to a trade-off of the reduction of the number of excitations and their energy spread.
For Lorentzian pulses injecting an integer amount of particles on the scatterer, the dependence of the energy noise on the number of emitted particles is a measure of their dependent or independent emission. All discussed effects are expected to be  
observable via power fluctuations.

\section{Acknowledgments}
We thank Julien Gabelli for insightful discussions. Financial support from the Ministry of Innovation, NRW, from the Excellence Initiative of the German Federal and State Governments, as well as from the the Swedish Research Council (JS), the Alexander von Humboldt Foundation (FH), and CONICET, Argentina (FB) are gratefully acknowledged.

\appendix
\section{Scattering theory for time-dependently driven systems}\label{app:PAscattering}

In this appendix, we briefly recall the main aspects of the scattering theory of photon-assisted transport developed by B\"uttiker and coworkers, which underly the investigations performed in this paper.\cite{Buttiker93,Pretre96,Pedersen98,Rychkov05}   

We consider a multiterminal conductor with a periodic potential $V_\alpha(t)$ applied to each of its contacts $\alpha$.
It turns out to be convenient to divide the conductor into different subsections, namely the reservoirs, the scatterer, and perfect ballistic leads connecting the scatterer to the reservoirs. 

The reservoirs are typically good metallic contacts with efficient screening properties. Therefore, as long as the frequency of the driving is sufficiently slow, deep inside each reservoir $\alpha$, a change of  the potential ${V}_{\alpha}(t)$ applied to that contact goes along with an equal shift of the band bottom, guaranteeing that  local electrostatic equilibrium is maintained at every time $t$.\cite{Buttiker93,Pedersen98}  As a consequence, the Fermi level in the reservoirs, and therefore the occupation of each state with momentum $k$,  {\em does  not} depend on the oscillating part of  ${V}_{\alpha}(t)$.

In contrast, close to the scatterer and in the leads the effective potential ``seen" by the charge carriers has in general a non-trivial dependence on time due to the charge injected into the conductor and the one induced in the neighbouring gates in order to preserve charge neutrality of the whole device.\cite{Buttiker93}  This effective potential should in general be evaluated self-consistently.~\cite{Buttiker93prl,Pretre96,Pedersen98} Here, however,  following also Refs.~\onlinecite{Rychkov05,Vanevic12,Dubois13},  we do not attempt a self-consistent treatment of the potential seen by the charge carriers and we simply assume that the time-dependent part of the driving voltages ${V}_{\alpha}(t)$ vanishes in the leads as we approach the scattering region. Moreover, we assume ${V}_{\alpha}(t)$ to vary slowly with respect to the Fermi wavelength,  such as to not induce any additional scattering.  Although, this assumption on the spatial dependence of  ${V}_{\alpha}(t)$ might seem rather artificial, it has  turned out to be an appropriate model to describe various experiments  on charge noise in ac driven coherent conductors.\cite{Schoelkopf98,Reydellet03,Dubois13nat,Gabelli13,Gasse13}

Since we assume the ac part of the driving to vanish close to the scatterer, the properties of the latter can be described by the time-independent (elastic) scattering matrix ${\bf s}_{\alpha \beta}(E)$ introduced in Sec.\ref{sec:model}.\cite{Pretre96,Pedersen98}  As mentioned in that section, the crucial point in order to be able to evaluate observables such as the average current or the noise is to determine quantum statistical averages of the form  ${\langle \hat{a}_{\alpha n}^{\dagger}(E)\hat{a}_{\beta m}(E')\rangle}$ or ${\langle \hat{a}_{\alpha n}^{\dagger}(E)\hat{a}_{\beta m}(E')\hat{a}_{\gamma k}^{\dagger}(E'')\hat{a}_{\delta \ell}(E''') \rangle}$, where the operator $\hat{a}_{\alpha n}(E)$ annihilates an electron in channel $n$ in lead $\alpha$ that impinges on the scatterer with energy $E$. This requires relating the operators $\hat{a}_{\alpha m}(E)$ for electrons close to the scatterer to those, which we call  $\hat{a}'_{\alpha m}(E)$, that correspond to the annihilation of electrons just injected by the reservoirs. Since the latter remain in electrostatic equilibrium at every time $t$, the statistical averages of the operators $\hat{a}'$ are simply equilibrium averages with respect to the local Fermi level $E_{\alpha,\mathrm{F}}=e\bar{V}_{\alpha}$, i.e.  ${\langle \hat{a}_{\alpha n}'^{\dagger}(E)\hat{a}'_{\beta m}(E')\rangle}=\delta_{\alpha\beta}\delta_{nm}\delta(E-E')f(E-E_{\alpha,{\rm F}})$.

To determine the relation between $\hat{a}_{\alpha m}(E)$ and $\hat{a}'_{\alpha m}(E)$ we observe\cite{Pretre96,Pedersen98}  that close to the scatterer,  where there is no oscillating potential, incoming particles in channel $m$ in lead $\alpha$ are described by  wave functions of the form $\psi_{\alpha m}(q)=\chi_{\alpha m}({\bf r}_{\perp})e^{iq x_{\alpha}}e^{-i\varepsilon_{q\alpha m}t}$, where $x_{\alpha}$ is the coordinate along the direction of propagation, $q$ is the longitudinal momentum and $\chi_{\alpha m}({\bf r}_{\perp})$ is eigenfunction of the $m$th transverse eigenmode. In the portion of lead $\alpha$ close to the reservoir, where the oscillating potential, $V_{\alpha}(t)$, is finite, the wave-function is instead given by ${\psi_{\alpha m}'(q)=\psi_{\alpha m}(q)e^{-i\phi_{\alpha}(t)}}$, where 
\begin{equation}\label{eq:phi}
\phi_{\alpha}(t)=\frac{e}{\hbar}\int_{0}^{t}dt'\,[V_{\alpha}(t')-\bar{V}_{\alpha}],
\end{equation}
is an additional time-dependent phase due to the oscillating part of the potential. Matching the wave functions in the regions with and without modulation potential, and changing to a description in second quantization, one finds the required relation  between operators $\hat{a}_{\alpha m}(E)$ and $\hat{a}_{\alpha m}(E)$:~\cite{Pedersen98}
\begin{equation}
\hat{a}_{\alpha m}(E)=\sum_{k=-\infty}^{+\infty}c_{\alpha k}\hat{a}'_{\alpha m}(E-k \hbar\Omega),
\label{app:aaprime}
\end{equation}
The coefficients $c_{\alpha k}$ are the Fourier components of $e^{-i\phi_{\alpha}(t)}$, as   defined in Eq.~\eqref{eq:ck}. Equation \eqref{app:aaprime} expresses the fact that 
the time-dependent part of the driving applied to a coherent conductor leads to a spread in energy of the wave-function, such that each state with energy $E$ impinging on the scattering region corresponds to  a superposition of states with different energies in the reservoirs.

The relationship between the operators $\hat{\bf b}_{\alpha}$ and $\hat{\bf a}'_{\alpha}$ for outgoing and incoming particles in the reservoirs can be synthetically expressed in terms of a Floquet scattering matrix $\boldsymbol{\mathcal{S}}$,
\begin{equation}
\hat{\bf b}_{\alpha}(E)=\sum_{\beta}\boldsymbol{\mathcal{S}}_{\alpha \beta}(E,E')\hat{\bf a'}_{\beta}(E'),
\end{equation}
with ${\boldsymbol{\mathcal{S}}_{\alpha\beta}(E,E')={\bf s}_{\alpha\beta}(E)\sum_{\ell=-\infty}^{\infty}c_{\beta \ell}\delta(E-E'-\ell\hbar\Omega)}$.
The matrix elements $\mathcal{S}_{\alpha\beta,mn}(E,E')$ represent the probability amplitude that an electron with energy $E'$ incoming from reservoir $\beta$ in channel $n$ is ejected at energy $E$ and in channel $m$ in contact $\alpha$. Note that since the driving is periodic, scattering can only occur between states that differ in energy by a multiple of the driving frequency.

\bibliographystyle{apsrev4-1}
%

\end{document}